\documentclass[aps, prl,superscriptaddress, twocolumn,showpacs,preprintnumbers,amsmath,amssymb]{revtex4}
\usepackage{graphicx}
\usepackage{bm}
\usepackage{amssymb, latexsym, amsmath, textcomp}
\usepackage{hyperref}

\begin{document}

\title{Spontaneous spin ordering of Dirac spin liquid in a magnetic field}

\author{Ying Ran}
\affiliation{Department of Physics, University of California, Berkeley, California 94720}

\author{Wing-Ho Ko}
\author{Patrick A. Lee}
\author{Xiao-Gang Wen}
\affiliation{Department of Physics, Massachusetts
Institute of Technology, Cambridge, Massachusetts 02139}
\date{\today}
\begin{abstract}
The Dirac spin liquid was proposed to be the ground state of the spin-1/2
Kagome antiferromagnets. In a magnetic field $B$, we show that the state with
Fermi pocket is unstable to the Landau level (LL) state. The LL state breaks
the spin rotation around the axis of the magnetic field.
We find that the LL state has an in-plane 120$^{\circ}$ $q=0$ magnetization $M$ which scales with the external field $M\sim B^{\alpha}$, where $\alpha$ is an intrinsic calculable universal number of the Dirac spin liquid. We discuss the related experimental
implications which can be used to detect the possible Dirac spin liquid phase in
Herbertsmithite ZnCu$_3$(OH)$_6$Cl$_2$.
\end{abstract}
 \maketitle

Spin liquids (SL), defined as the ground states of spin systems with half
integer spins per unit cell which does not order magnetically and/or break
translation symmetry, are believed to contain fundamentally new physics beyond
the Landau's symmetry breaking characterization of phases. After years of
search, a promising candidate finally emerged in the spin-1/2 Kagome system
Herbertsmithite
ZnCu$_3$(OH)$_6$Cl$_2$\cite{helton-2007-98,mendels-2007-98,ofer-2006}. Despite
an antiferromagnetic exchange $J\approx
170\sim190$K\cite{misguich-2007,rigol:184403}, the system does not order down to
$50$mK\cite{helton-2007-98,mendels-2007-98,ofer-2006}. Theoretically the
Dirac-SL is proposed\cite{ran-2007-98} as the ground state of the nearest
neighbor Heisenberg antiferromagnetic model on Kagome lattice. Unfortunately
the spin susceptibility $\chi_s$ is consistent with $\sim 4$\% magnetic
impurities\cite{misguich-2007}, and possible Dzyaloshinskii-Moriya interaction
has also been proposed\cite{rigol:184403}. These may explain the hump of the
specific heat $C$ around $2$K and also obscure the $C\propto T^2$ and
$\chi_s\propto T$ behaviors predicted by the Dirac-SL. In this paper we
propose a unique signature of the Dirac-SL in an
external magnetic field, which can be used to detect the Dirac-SL phase in
experiment and numerical simulations.

\begin{figure}
\begin{center}
\includegraphics[width=0.2\textwidth]{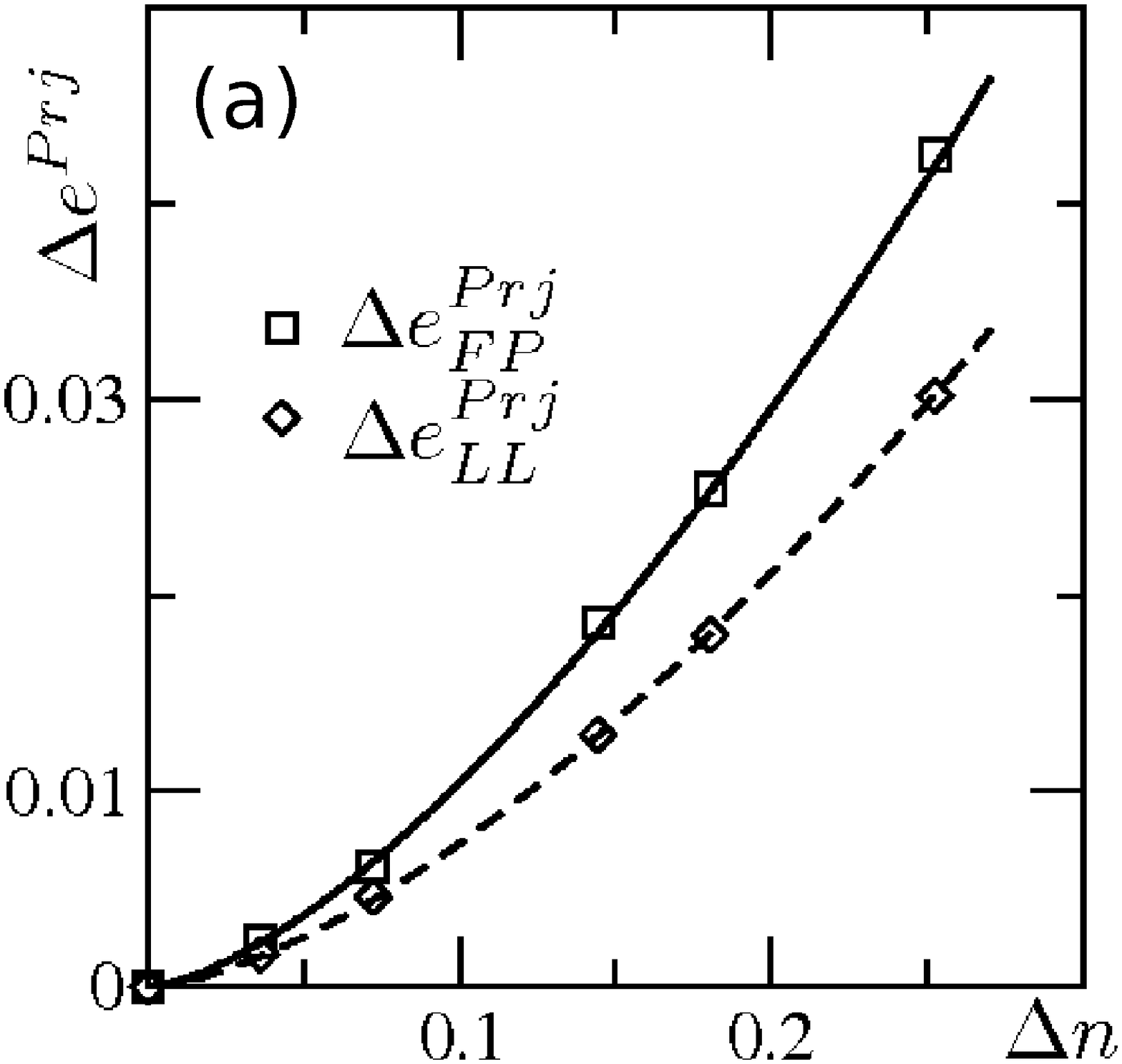}\hfil\includegraphics[width=0.22\textwidth]{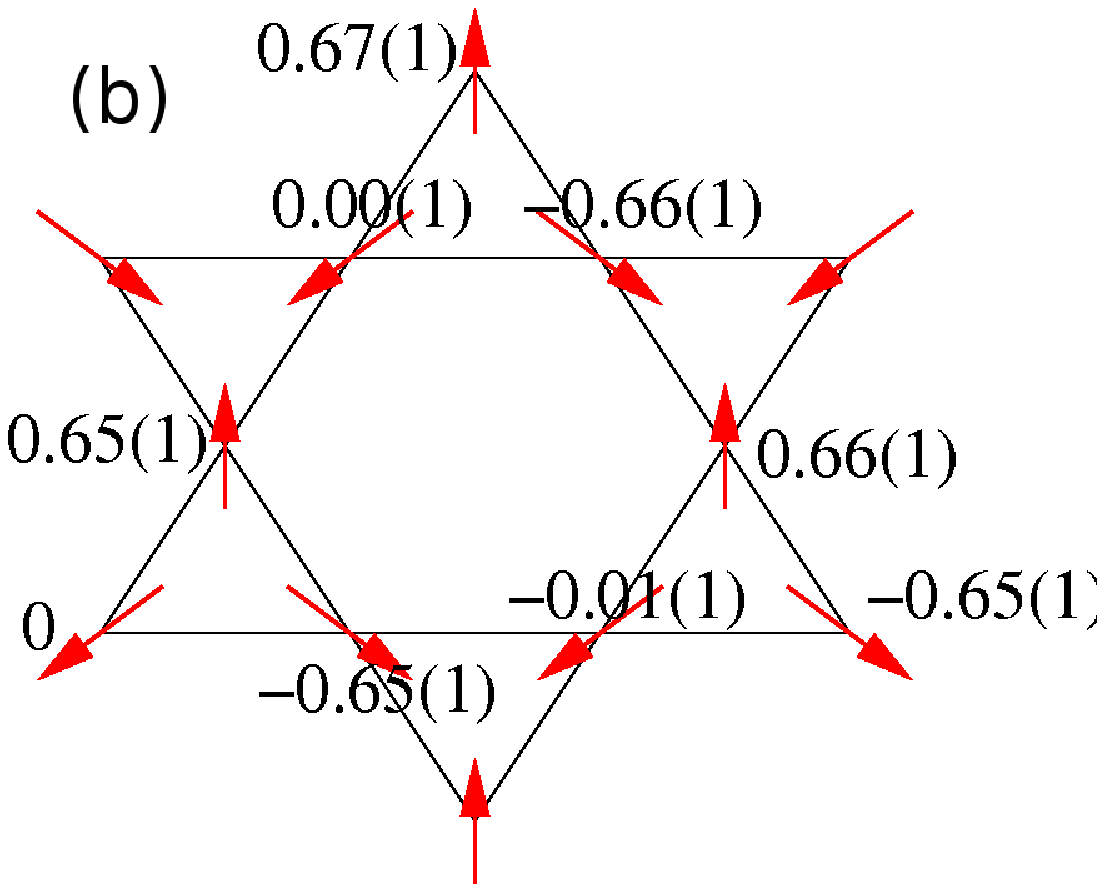}
\end{center}
\caption{(color online) (a) We convert the data in Table \ref{LL_PJ} into energy density $\Delta
e$ and spin imbalance density $\Delta n$ (units defined in text), and plot them together with the fitted scaling form $\Delta e=a\Delta n^{3/2}+b\Delta n^2$. (b): The XY spin order in the LL state. We also present the relative phase of $S_i^+$ operator computed by projected landau level state in unit of $\pi$ (see Eq.(\ref{relative-phase})). Note that we fix the left-bottom corner site as the site $i$ in Eq.(\ref{relative-phase}), and scan the site $j$ for the nine sites depicted.}\label{MF_FIT}
\end{figure}

The Dirac-SL experiences no orbital effect in the external magnetic field because it is an insulator. The Zeeman coupling $g\mu_B \vec B\cdot \sum_i\vec S_i$ polarizes spin along the field direction, breaks time reversal and breaks $SU(2)$ spin rotation down to $U(1)$. Let us denote the direction of $\vec B$ as $z$-direction (note that the xy-plane does not have to be the plane of the two-dimensional Kagome system). We find that the Dirac-SL will spontaneously break the remaining $S_z$-$U(1)$ and form a staggered magnetization $M$ (see Fig.\ref{MF_FIT}(b)) in the xy-plane which scales with $B$: $M\sim B^{\alpha}$. The positive exponent $\alpha$ is an intrinsic universal number of the Dirac-SL phase and in principle calculable.

This unique signature of the Dirac-SL can be compared with a regular co-planer antiferromagnetic (AF) ordered phase. In a small magnetic field $B$, the magnetization of the regular AF phase would rotate into the xy-plane to maximize the susceptibility along the $B$ direction. As $B$ is tuned to zero, the in-plane magnetization remains finite. In contrast, for the Dirac-SL we predict that $M$ vanishes as $B^\alpha$. This suggests that the Dirac-SL can be viewed as an AF phase whose long-range AF order has been destroyed by the quantum fluctuations.

We begin with writing down the the low energy theory of the Dirac-SL\cite{ran-2007-98}, which includes four flavors of fermions
coupled with compact $U(1)$ gauge field in 2+1 dimension (QED$_3$):
\begin{align}
\label{effS}
S=&\int dx^3 \big[
\frac{1}{g^2}(\varepsilon_{\lambda\mu\nu}\partial_{\mu}a_{\nu})^2
+\sum_{\sigma}\bar{\psi}_{+\sigma}\left(
\partial_{\mu}-ia_\mu\right)\tau_{\mu}\psi_{+\sigma}\notag\\
&+\sum_{\sigma}\bar{\psi}_{-\sigma}\left(
\partial_{\mu}-ia_\mu\right)\tau_{\mu}\psi_{-\sigma}\big] + \cdots ,
\end{align}
where the two-component fermionic Dirac spinon fields are denoted by
$\psi_{\pm \sigma}$, where $\pm$ label the two inequivalent nodes and $\sigma$
the up/down spins.

The spinon carries spin-1/2 and charge-0, and couples
to the external magnetic field only by the Zeeman effect. In the presence of a
magnetic field, the simplest guess is that the Dirac points will change into
Fermi pockets due to Zeeman splitting. We will call this state Fermi pocket
state (FP). But, can there be other states whose energy may be lower?

\begin{figure}
\includegraphics[width=0.2\textwidth]{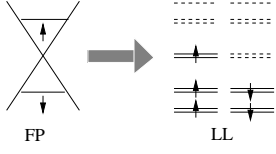}
\caption{The
The spinons is gapless in FP state 
with an electron-like Fermi pocket for spin up spinons and
hole-like pocket for spin down spinons. 
The spinons are fully gapped in the LL state. 
Each Landau level is doubly degenerate due to two inequivalent
Dirac nodes.}
\label{FP_LL}
\end{figure}

The FP state has many gapless spinon excitations near the Fermi pockets and in general not energetically favorable. One natural way to
gap out the Fermi pocket state is to induce (internal) gauge fluxes and
develop Landau levels. Here we take advantage of the appearance of zero energy
Landau levels when Dirac particles are subject to a magnetic field, as is well
known in the recent studies of
graphene\cite{novoselov-2005-438,zhang-2005-438}. If the gauge flux is
adjusted in such a way that the zero-energy Landau level is fully filled for
up-spin, and fully empty for down-spin, then the spinons are fully gapped
(Fig.\ref{FP_LL}).  We will call this state Landau level state(LL).

In the following we show that LL state has lower energy than FP state. We first compare the energies of LL state and FP state at the mean-field level (ignoring the gauge
fluctuation) with fixed $S_z$
polarization. 
We set aside the Zeeman energy which is common between the two states. The
density of spin imbalance is:
\begin{align}
\label{n_def}
\Delta n=\frac{\Delta N}{A}=
\frac{N_{\uparrow}-N_{\downarrow}}{A}
=4\cdot\frac{1}{4\pi}\left(\frac{\mu_B B}{v_F}\right)^2 ,
\end{align}
and the mean-field energy density is found to be
\begin{align}
\label{FP_MF}
\Delta e_{MF}^{FP}&=\frac{\Delta
E_{MF}^{FP}}{A}=\frac{2\sqrt{\pi}v_F}{3}(\Delta
n)^{\frac{3}{2}},
\end{align}
where $N_{\uparrow}$ and $N_{\downarrow}$ are the number of up and down spins,
$\Delta E$ is the energy increase compared to Dirac point state, $A$ is the
system area, $B$ is the magnetic field, and $v_F$ is the mean-field Fermi
velocity. The factor $4$ in Eq.(\ref{n_def}) is from the fact that there are
two spin degeneracy and two nodal degeneracy. We choose our units such that
$\hbar=1$.

To make sure the LL state have the same spin imbalance as the FP state, each
Landau level should contain $\Delta n A/2$ states, i.e. the induced gauge
magnetic field $b$ satisfies $\Delta n=2\cdot \frac{b}{2\pi}$, where the
$2\pi$ is gauge flux quantum. Since at the mean-field level the internal gauge field costs no energy,
the energy of the mean-field LL state is just the difference between the sum
of the energies of all negative landau levels, and the filled Fermi sea:
\begin{align}
\label{LL_MF}
\Delta e_{MF}^{LL}&=4v_F\left(-\frac{b}{2\pi}\sum_{n=1}^{\infty}\sqrt{2 n b}
- \int\frac{dk^2}{(2\pi)^2}(- k) \right)\notag\\
&=\frac{\zeta(\frac{3}{2})b^{3/2}v_F}{\sqrt{2}\pi^2}
=\frac{\zeta(\frac{3}{2})\Delta
n^{3/2}v_F}{\sqrt{2\pi}}.
\end{align}
The same result was obtained in Ref
\cite{PhysRevD.29.2366,PhysRevLett.64.1166}. From Eq.(\ref{FP_MF},\ref{LL_MF}) we have $\frac{\Delta
e_{MF}^{FP}}{\Delta e_{MF}^{LL}}=\frac{2\sqrt{2} \pi}{3\zeta(3/2)}\doteq1.134$,
i.e., the LL state has lower mean field energy than the FP state.

We may also ask whether in graphene a spontaneous magnetic field may be
generated by this mechanism, if electrons or holes are introduced by gating.
The answer is negative and the difference is that the physical electromagnetic
field has an energy cost of $f_{\mu\nu}^2$. Furthermore the field is three
dimensional and the magnetic field must form closed lines. The magnetic flux
would be divided into domains of size $R$ with opposite signs. The size $R$
can be estimated by minimizing the total energy, which is the sum of the
magnetic field energy $\sim B^2R^3$ and the Landau level energy gain $\sim
-v_F\sqrt{\hbar e B/c}\; B R^2/\Phi_0$ where $\Phi_0=\frac{hc}{e}$ is the flux
quantum. The optimal $R_*$ is found to satisfy
$\frac{BR_*^{2}}{\Phi_0}\sim(\alpha\frac{v_F}{c})^2$, where
$\alpha=\frac{e^2}{\hbar c}=\frac{1}{137}$ is the fine structure constant.
This means that the flux through the entire domain is $\sim 10^{-7}\Phi_0$ ,
which is not self consistent with the condition that there is at least one
flux quantum through the domain to support Landau levels.

To include the effect of gauge fluctuations and to go beyond the mean field
theory, we will use the Gutzwiller projected wavefunction to calculate the
energies of LL and FP states.  To obtain the Gutzwiller projected wavefunction we first
write down the mean-field Hamiltonian on lattice:
$H_{mean}=\sum_{ij}\chi_{ij}f_{i\sigma}^{\dagger}f_{j\sigma}$, where
$f_{i\sigma}$ are the fermionic spinons. The mean-field ground state $\vert
\Psi_{mean}(\chi_{ij}) \rangle$ with mean-field parameters $\chi_{ij}$ is a
spin singlet. The projected wavefunction $\vert
\Psi_{prj}(\chi_{ij}) \rangle=P \vert \Psi_{mean}(\chi_{ij}) \rangle$
removes the unphysical states and becomes a spin state; here
$P=\prod_i(1-n_{i\uparrow}n_{i\downarrow})$ is the projection operator
ensuring one fermion per site. The physical observables can be measured on $\vert \Psi_{prj}(\chi_{ij})
\rangle$ by a Monte Carlo approach\cite{3381199}. The Dirac-SL is
characterized by $\chi_{ij}$ such that $|\chi_{ij}|=\chi$ is a bond
independent constant and there are $\pi$ fluxes through the hexagonal
plaquettes and $0$ fluxes through the triangular plaquettes\cite{ran-2007-98}.
Note that the projected Dirac-SL state has no tunable parameter since $\chi$ only gives the wavefunction an overall factor. On a 16x16 unit cell lattice, the energetics of the FP and LL states are given in
Table \ref{LL_PJ}.
\begin{table}
\begin{tabular}{|l|l|l|l|}
\hline
$\Delta
N$&0&8&16\\
\hline
$E^{FP}_{\textrm{Prj}}$&-0.42865(2)&-0.42798(2)&-0.42688(2)\\
\hline
$E^{LL}_{\textrm{Prj}}$&-0.42865(2)&-0.42817(2)&-0.42732(2)\\
\hline\hline
$\Delta
N$&32&40&56\\
\hline
$E^{FP}_{\textrm{Prj}}$&-0.42327(2)&-0.42131(2)&-0.41638(2)\\
\hline
$E^{LL}_{\textrm{Prj}}$&-0.42493(2)&-0.42346(2)&-0.41994(2)\\
\hline
\end{tabular}
\caption{The per-site energies of the projected FP and LL states
 in unit of Heisenberg coupling $J$ on a 16x16 unit cell lattice with periodic-antiperiodic boundary conditions. $\Delta N$ is the spin imbalance. The FP states has no total gauge flux through the sample, while the LL state has $\frac{\Delta N}{2}$ flux quanta through the sample which is uniformly distributed in the triangular and hexagonal plaquettes.}\label{LL_PJ}
\end{table}

Is the $\Delta e\sim\Delta n^{3/2}$ law still
valid after projection? The answer is positive because 
the $S_z$ and energy are conserved and have no anomalous dimension. The
energies in Table \ref{LL_PJ} can be thus fitted (Fig.\ref{MF_FIT}(a)) as
\begin{align}
\Delta e^{Prj}_{FP}&=0.33(2)\Delta n^{3/2}+0.00(4)\Delta n^2\label{FP_PJ_LT},
\\
\Delta e^{Prj}_{LL}&=0.223(6)\Delta n^{3/2}+0.03(1)\Delta n^2
\label{LL_PJ_LT},
\end{align}
where the units are chosen such that the unit cell spacing $a=1$ and $J=1$ and we include the first order correction to scaling, the $\Delta n^2$ term.  From the coefficient
$0.33(2)$ in Eq.(\ref{FP_PJ_LT}) we are able to read off the effective fermi velocity
$v^*_F=\frac{3\cdot 0.33(2)}{2\sqrt{\pi}}=0.28(2)\frac{Ja}{\hbar}$ by fitting
to the free fermion result Eq.(\ref{FP_MF}). $v^*_F$ is almost twice
the mean-field value $v_F=\frac{a\chi}{\sqrt{2}\hbar}$ with $\chi=0.221J$\cite{PhysRevB.63.014413}. For
Herbertsmithite this means $v^*_F=4.9\cdot 10^3$m/s assuming $J=200$K. This is
very close to the fermi velocity we found by a projected band structure
study \cite{hermele:224413}, where only one particle-hole excitation is considered. The closeness of the two results imply that the
gauge interactions between many particle-hole excitations may only give small
corrections for energetics.

From Eq. (\ref{FP_PJ_LT},\ref{LL_PJ_LT}) we see that the LL state has a lower
energy than the FP state.  Therefore the LL state may be the true ground state
in the presence of a magnetic field. Due the presence of internal gauge flux,
one can easily see that the LL state breaks parity (mirror reflection). Surprisingly, it turns out that the LL state breaks the $S_z$ spin rotation as well. To understand this let
us consider the low energy collective excitations in the LL state.  Since the
spinons $\psi$ are gapped due to the Landau levels, the gauge field $a_\mu$ is the only low
energy excitation. We note that the total Hall conductance of the
spin-up spinons is $+1$ and the total Hall conductance for the spin-down
spinons is $-1$.  Thus the total Hall conductance is zero and there is no
Chern-Simons term for the gauge field $a_\mu$.  As a result, the dynamics of
the gauge field is controlled by the Maxwell term $f_{\mu\nu}^2$, which arise from integrating out high-energy fermions, and gives rise to a linear dispersed gapless photon mode.

To understand the meaning of the gapless mode, we note that due to the
non-zero Hall conductances $\pm 1$ for the spin-up and spin-down spinons, a
flux quantum of the  gauge field $a_\mu$ will induce one spin-up spinon and
minus one spin-down spinon.  Thus a flux quantum carries a spin quantum number
$S_z=1$.  The ``magnetic field'' of $a_\mu$ is nothing but the $S_z$ density.
Therefore, the linear gapless mode of $a_\mu$ is actually the density
fluctuations of $S_z$. The appearance of a linear gapless mode of a conserved
density as the only low lying excitation implies that the corresponding
symmetry is spontaneously broken in the ground state.\cite{PhysRevB.41.240,PhysRevLett.69.1811} Thus
the LL state contains an XY order: $\langle S^+ \rangle \neq 0$ and the
gapless gauge field $a_\mu$ is the Goldstone boson mode of the $U(1)$ symmetry
breaking.  The ``electric charge'' of $a_\mu$ gauge field corresponds to the
vortex in the XY ordered phase. This duality \cite{PhysRevLett.47.1556,PhysRevB.39.2756} between the spin degrees of freedom and the gauge degrees of freedom allows us to translate the physics in the two languages back and forth.

For example let us consider the monopole insertion operator $V^{\dagger}$ which inserts a $2\pi$ gauge flux. $V^{\dagger}$ flips one spin as mentioned before and can be expanded as $V^{\dagger}\sim \sum_{i} e^{i\theta_i} S_i^{+}$, where the summation is restricted to the area where the $2\pi$ flux is inserted. Because $V^{\dagger}$ has long-range correlation in the Coulomb phase, by the above duality, the spin correlation $\langle e^{i\theta_i} S_i^{+},e^{-i\theta_j} S_j^{-}\rangle$ is also long ranged.

To study what kind of the XY spin order that the LL state has, we have
calculated spin-spin correlation function $\langle  S_{x,i} S_{x,j}+ S_{y,i} S_{y,j} \rangle$ of the projected LL state and found the XY ordered pattern as shown in Fig.\ref{MF_FIT}(b). This order pattern is referred as the $q=0$ magnetic order in literatures\cite{PhysRevB.45.2899}. To confirm this is indeed the order pattern we directly compute the relative phases of $S_i^{+}$ operators in the monopole insertion operator $V^{\dagger}$:
\begin{align}
\theta_j-\theta_i=\arg\Big(\frac{
\langle n+1 \text{ flux quanta}|P S^+_jP |n \text{ flux quanta} \rangle }{
\langle n+1 \text{ flux quanta}|P S^+_iP |n \text{ flux quanta} \rangle }\Big) ,
\label{relative-phase}
\end{align}
where $ |n \text{ flux quanta} \rangle $ is a LL state with $n$ flux quanta and $P$ is
the projection operator. To understand this formula one should note that $P|n+1 \text{ flux quanta} \rangle=V^{\dagger}P|n \text{ flux quanta} \rangle$, where $V^{\dagger}$ inserts an extra $2\pi$ flux uniformly while flipping one spin, and thus $V^{\dagger}\sim\sum_{i}e^{i\theta_i}S_i^+$ with summation over all sites. These relative phases can be computed analytically by choosing proper gauge (i.e., gauge that respects physical symmetries). To get the answer quickly we perform the Monte Carlo calculation on a finite lattice. For a 6 by 6 unit cell sample on torus with $n=4$, we compute this relative phase for 9 sites within the three adjacent unit cells and the result is given in Fig. \ref{MF_FIT}(b). We find the relative phases are $0,\pm\frac{2\pi}{3}$ alternating between the three sublattices and this is exactly the $120^{\circ}$ $q=0$ order pattern.\footnote{We actually find that this relative phase is independent of $n$. Even when $n=0$, namely the pure $U(1)$-Dirac SL state, the relative phase is the same as the $n=4$ data that we presented up to statistical error. This indicates that the relative phase which we computed is an intrinsic property of the $U(1)$-Dirac SL, i.e., they are the spin-1 monopole quantum numbers. The Eq.(\ref{relative-phase}) is a systematic way to compute monopole quantum numbers in any $U(1)$-Dirac SL. A more detailed discussion on this method is presented as an appendix in Ref.\cite{ran-af-sc-2008}. The monopole quantum numbers obtained in this method (see Fig. \ref{MF_FIT}(b)) are consistent with the results obtained by an independent study on the Dirac SL in zero magnetic field based on symmetry group analysis\cite{hermele:224413}. }

Because $V^{\dagger}$ is in fact the magnetization operator, we conclude that the in-plane staggered magnetization $M\sim B^{\alpha}$ scales with the external magnetic field, where the exponent $\alpha$ is the scaling dimension of the monopole operator at the Dirac-SL fixed point, because $B$ is dimension-1. This exponent is computable by numerical simulations of QED$_3$ or the field theory techniques such as $1/N$ expansions, and subject to further study.

So far we show that the FP state is unstable towards the $S_z$-broken LL state by numerical arguments. In the following we present an analytical argument that the $S_z$ symmetry is broken by studying the low energy effective theory. The FP state is characterized by a electron-like Fermi pocket of spin up
spinons $f_{k\uparrow}$ and a hole-like Fermi pocket of spin down spinons
$f_{k\downarrow}$. After doing a particle-hole transformation on the spin
down Fermi pocket only $f_{k\downarrow}\rightarrow h_{-k\uparrow}$ we have $f_{k\uparrow}$ and $h_{-k\uparrow}$ carry same spin but opposite gauge
charges. The ``Coulomb'' attraction between the two particles will cause a
pairing instability $\langle
f_{k\uparrow}^{\dagger}h_{-k\uparrow}^{\dagger}\rangle=\langle
f_{k\uparrow}^{\dagger}f_{k\downarrow}\rangle\neq 0$. This is a triplet excitonic insulator and this has been discussed in the context of graphene\cite{khveshchenko:115104,aleiner:195415}. Besides Coulomb
attraction, there is also an Amperean attraction\cite{lee:067006} between the
currents of $f_{k\uparrow}^{\dagger}$ and $h_{-k\uparrow}^{\dagger}$
excitations. The ``Coulomb'' attraction and the Amperean attraction are cooperating for the
same condensation. Note that $f_{k\uparrow}^{\dagger}f_{k\downarrow}$ is a
$S_z$ spin-1 object. 

Finally we discuss the consequences of this spontaneous spin ordering in experiment. First the in-plane $q=0$ magnetization pattern and its scaling law $M\sim B^{\alpha}$ is observable by neutron scattering. Second since the ground state breaks the parity and $S_z$ rotation symmetry, it is
separated from the high temperature paramagnetic phase by at least one
\emph{finite} temperature phase transition. This transition can be first order or continuous. If the transition is a continuous one that restores the $S_z$-$U(1)$ symmetry, we expect it to be the Kosterlitz-Thouless universality. Here we simply estimate the transition temperature $T_c\sim\frac{\mu_B B}{k_B }$ when $\mu_B B\lesssim \chi^{Prj}$. This is because the external magnetic field $B$ is the only energy scale if it is much smaller than the spinon band width $\chi^{Prj}$. Since these are intrinsic properties
of the Dirac-SL in magnetic field, they can be used to experimentally detect the possible Dirac-SL ground state in Herbertsmithite and other materials where a Dirac-SL may be realized. This XY spin ordering can also serve a way to detect Dirac-SL in numerical studies of the Kagome lattice Heisenberg model such as exact diagonalization.

There are other proposals for the possible non-magnetic Valence Bond Solid (VBS) ground states in the spin-$1/2$ Heisenberg model on Kagome lattice\cite{PhysRevB.68.214415,singh:180407}. If one of those VBS states is the ground state of Herbertsmithite, it will surely not break $S_z$ symmetry in a small magnetic field because the VBS phases are fully gapped. The $S_z$ symmetry can be broken at an external magnetic field larger than the spin gap due to the triplon condensation. However the XY magnetic order generated in this fashion is unlikely to be the $q=0$ pattern because the VBS orders itself breaks translation and we expect their XY orders also breaks translation and have a large unit cell. Therefore the spontaneous spin ordering we present in this work can be used to differentiate the $U(1)$-Dirac SL from VBS states experimentally.

It is likely that in the current Herbertsmithite compound there is an energy scale below which the significant amount of impurities and/or the Dzyaloshinskii-Moriya(DM) interaction start to play an important role. Recent experiments estimate the strength of the DM interaction to be $\sim 15$K\cite{zorko:026405}. In a strong magnetic field $\sim30$~tesla we expect that it is possible to suppress their effects and reveal the intrinsic property of the Dirac-SL. 

We thank M. Hermele, T. Senthil and F.-C. Zhang for helpful discussions and T.
M. Rice for raising the issues concerning graphene. This research is supported
by NSF grant DMR-0706078 and DMR-0804040.

\bibliographystyle{apsrev}
\bibliography{/home/ranying/downloads/reference/simplifiedying}

\end{document}